\documentclass[prl,aps,showpacs,
twocolumn,
]{revtex4}

\usepackage{graphicx}
\usepackage{amsmath}
\usepackage{amssymb}
\usepackage{hyperref} 

\newcommand{\be}{\begin{equation}}
\newcommand{\ee}{\end{equation}}
\newcommand{\bea}{\begin{eqnarray}}
\newcommand{\eea}{\end{eqnarray}}

\newcommand{\0}{\over }

\begin{document}

\title{Violation of the Holographic Viscosity Bound\\ in a Strongly Coupled Anisotropic Plasma}

\preprint{TUW-11-28}

\author{Anton Rebhan}
\author{Dominik Steineder}
\affiliation{Institut f\"ur Theoretische Physik, Technische Universit\"at Wien,
        Wiedner Hauptstrasse 8-10, A-1040 Vienna, Austria}

\date{\today}

\begin{abstract}
We study the conductivity and shear viscosity tensors of a strongly coupled $\mathcal N=4$ super-Yang-Mills plasma which is kept anisotropic by a $\theta$ parameter that depends linearly on one of the spatial dimensions. Its holographic dual is given by an anisotropic axion-dilaton-gravity background and has recently been proposed by Mateos and Trancanelli as a model for the pre-equilibrium stage of quark-gluon plasma in heavy-ion collisions. By applying the membrane paradigm which we also check by numerical evaluation of Kubo formula and lowest lying quasinormal modes, we find that the shear viscosity purely transverse to the direction of anisotropy saturates the holographic viscosity bound, whereas longitudinal shear viscosities are smaller, providing the first such example not involving higher-derivative theories of gravity and, more importantly, 
with fully known gauge-gravity correspondence. 
\end{abstract}
\pacs{11.25.Tq, 11.10Wx, 12.38.Mh}

\maketitle

{\it Introduction.} 
Hydrodynamic simulations of heavy-ion collisions suggest \cite{Romatschke:2007mq} that the produced quark-gluon plasma is behaving like an almost perfect fluid with a ratio of shear viscosity over entropy density not far from the famous result $\hbar/4\pi$ associated with the membrane paradigm of
black holes \cite{KPM:BHMP} and 
which holographic gauge-gravity duality maps to the corresponding quantity of maximally supersymmetric Yang-Mills theory in the limit of infinite color number and infinite 't Hooft coupling \cite{Aharony:1999ti,Policastro:2001yc}. This value has been conjectured to form the lower bound for any realistic matter \cite{Kovtun:2004de}. It was found to be saturated universally \cite{Buchel:2003tz,Iqbal:2008by} in dual theories involving an isotropic horizon described by Einstein gravity. 
Values above this bound are obtained when corrections due to finite 
coupling strength are included \cite{Buchel:2008ae}, but it has been shown that values violating the bound can 
arise in higher-derivate gravities \cite{Brigante:2007nu
}, although so far no complete gauge-gravity correspondence has been established for finite violations. 

The possibility of nonuniversal shear viscosity within two-derivative gravity has been considered before in gravity duals of noncommutative gauge theories \cite{Landsteiner:2007bd} and p-wave superfluids \cite{Natsuume:2010ky}, where shear modes in addition to the universal purely transverse one exist. For p-wave superfluids, nonuniversal values above the bound have been recently obtained in \cite{Erdmenger:2010xm
}.

Anisotropies are of particular importance to the nonequilibrium dynamics of the quark gluon plasma in heavy-ion collisions.
At weak coupling, they give rise to non-Abelian plasma instabilities \cite{Mrowczynski:1988dz
} and turbulent behavior \cite{Arnold:2005ef
} that has been argued to effectively reduce the shear viscosity \cite{Asakawa:2006tc}, which is parametrically large in a weakly coupled plasma \cite{Arnold:2003zc
}. More directly accessible experimental signatures such as anisotropic photon and dilepton emission \cite{Schenke:2006yp}, momentum broadening of jets \cite{Romatschke:2006bb
} as well as quarkonium dissociation \cite{Burnier:2009yu} have been studied in the hard-loop effective theory \cite{Mrowczynski:2004kv} for a weakly coupled anisotropic plasma where on sufficiently small time scales the anisotropy can be taken as stationary.

While holographic models of the evolution of strongly coupled quark-gluon plasma towards isotropy are of immediate interest and have been studied extensively \cite{
Hubeny:2010ry}, further insight may be obtained by models where the anisotropy is temporarily fixed. The first such attempt was presented by Janik and Witaszczyk in \cite{Janik:2008tc}, where the gravity dual involves a naked singularity. In \cite{Rebhan:2011ke} we have studied electromagnetic signatures of this model which are qualitatively similar to weak-coupling results at high frequencies, but which involve vanishing conductivities in the limit of zero frequency and absence of hydrodynamic behavior.

{\it Anisotropic axion-dilaton-gravity solution.}
Recently, Mateos and Trancanelli \cite{Mateos:2011ix,Mateos:2011tv} have adapted the spatially anisotropic duals of Lifshitz-like fixed points of \cite{Azeyanagi:2009pr} to a setup with AdS boundary conditions and finite temperature. This provides an anisotropic version of an $\mathcal N=4$ super-Yang-Mills plasma where the anisotropy is kept stationary by deforming the gauge theory by
\be
\delta S=\frac{1}{8\pi^2}\int\theta(z)\,\textnormal{Tr }F \wedge F
\ee
with
$\theta(z)=2\pi a z$ depending linearly on one of the spatial dimensions. The constant $a$ is related to the density of D7-branes (homogeneously distributed along $z$) that are dissolved in the bulk of the dual theory (they do not extend to the holographic boundary). The dual bulk theory is effectively 5-dimensional axion-dilaton gravity with negative cosmological constant and bulk action
\be\label{bulkaction}
S_{\rm bulk}=\frac{1}{2\kappa^2}\int_{\mathcal{M}}\!\!\sqrt{-g}\,\Bigl(R+12-\frac{(\partial \phi)^2}{2}-e^{2\phi}\frac{(\partial \chi)^2}{2}\Bigr)
\ee
where $\kappa^2=8\pi G=4\pi^2/N_c^2$ and axion $\chi=az$. Here we have dropped a factor $S^5$ of the ten-dimensional solution $\mathcal M\times S^5$ and set its radius $L=1$.
The remaining five-dimensional geometry which asymptotes to AdS$_5$ for $u\to0$ can be given by a line element of the form
\bea\label{ds2}
ds^2&=&\frac{e^{-\phi(u)/2}}{u^2}\Big(-\mathcal{F}(u)\mathcal{B}(u)\,dt^2+\frac{du^2}{\mathcal{F}(u)}\nonumber\\
&&\qquad+dx^2+dy^2+\mathcal{H}(u)\,dz^2\Big).
\eea
The functions $\mathcal{B},\mathcal{F}$ and $\mathcal{H}$ can be written in terms of the dilaton $\phi$, which itself has to satisfy a third-order nonlinear differential equation in $u$ \cite{Mateos:2011tv}. For nonvanishing anisotropy parameter $a$ there is an anisotropic horizon $\mathcal F(u_h)=0$ with $\mathcal H>1$, which is completely regular. This will allow us to apply the membrane paradigm along the lines of \cite{Iqbal:2008by} for studying conductivities and shear viscosities, both of which have two independent components at nonzero anisotropy. 

The thermodynamics of this setup and its instabilities are discussed in \cite{Mateos:2011tv}. Upon
holographic renormalization which brings in a reference scale $\mu$,
the stress tensor of the gauge theory has the form
\be
\langle T^{\mu\nu} \rangle = {\rm diag}(\epsilon,P_\perp,P_\perp,P_z),
\ee
with a conformal anomaly $\langle T^\mu_{\mu} \rangle \propto a^4$.
The entropy density is given by 
\be\label{entropy}
s=(\epsilon+P_\perp)/T
\ee
with
$T=|\mathcal F'(u_h)|\sqrt{\mathcal B_h}/4\pi$. Unlike the strictly positive $s$ which continues to be given
by a quarter of the horizon area over spatial volume, $\epsilon$ and
the pressure components depend separately on $T/\mu$ and $a/\mu$ and can
become negative in various regions of the phase diagram.
When $|a|$ is increased from the isotropic limit $a\!=\!0$, the pressure
anisotropy first becomes oblate ($P_z<P_\perp$), while prolate
anisotropies can be obtained for larger $|a|$.
For small $a$ there is a qualitative similarity to the plasma instabilities
in an anisotropic weakly coupled plasma in that the homogeneous phase is
unstable against filamentation of the distribution of the dissolved D7 branes along $z$
\cite{Mateos:2011tv}.
For sufficiently large $a$ the instabilities against filamentation however disappear, and also various components of the stress tensor (but not the entropy) can become negative.

\begin{figure}
\includegraphics[width=0.45\textwidth]{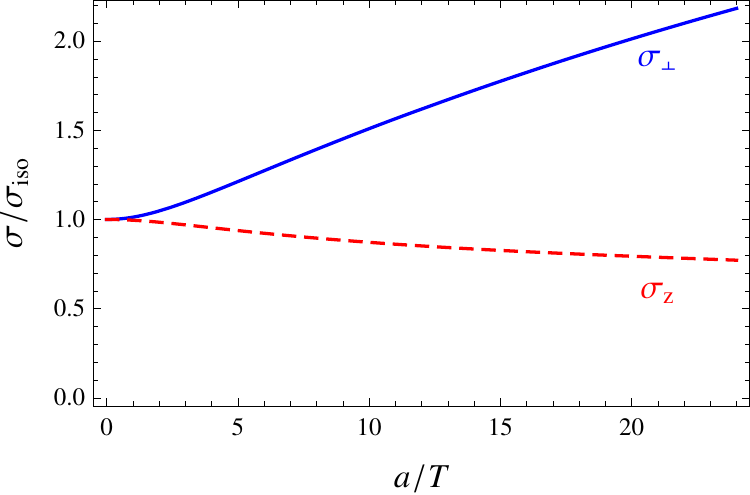}
\caption{DC conductivities along and transverse to the direction of anisotropy as a function of the anisotropy parameter $a/T$. 
\label{fig:sigmas}}
\end{figure}

{\it Conductivities.} Including a standard Maxwell action for a U(1) gauge field in the bulk with coupling constant $g_{\rm eff}$ leads to a conjugate momentum with respect to evolution in the coordinate $u$ that defines the current
\be
j^\mu=\frac{\partial \mathcal{L}}{\partial(\partial_u A_\mu)}=-\frac{\sqrt{-g}}{g_{\rm eff}^2}g^{uu}g^{\mu\nu}F_{u\nu}.
\ee
As in \cite{Iqbal:2008by}, regularity at the horizon in terms of the Eddington-Finkelstein coordinate $v$ for infalling observers defined by
\be
dv=dt-\sqrt{\frac{g_{uu}}{-g_{tt}}}du
\ee
relates $j^i$ linearly to the electric field $F_{ti}$, so that the horizon acts as a conductor. The anisotropic metric (\ref{ds2}) however leads to two different conductivities
\bea
\sigma_\perp&=&{g_{\rm eff}^{-2}}\sqrt{\gamma}\,g^{xx}|_{u_h}\\
\sigma_z&=&{g_{\rm eff}^{-2}}\sqrt{\gamma}\,g^{zz}|_{u_h}=\sigma_\perp/\mathcal H(u_h)\le \sigma_\perp
\eea
where $\gamma=g/g_{tt}g_{uu}$. In the limit of vanishing frequency $\omega$ and wave vector $\vec q$ we have $\partial_u j^\mu\to0$ which allows us to conclude that $\sigma_{z,\perp}$ give the DC conductivities of the boundary theory. The numerical results as a function of the anisotropy parameter $a/T$ are given in Fig.~\ref{fig:sigmas}.

Because $\sigma_{z,\perp}$ are determined by geometric data, they do not depend on $\mu$ and thus only on the ratio $a/T$. Whether a given value $a/T$ corresponds to oblate or prolate anisotropy however depends on $T/\mu$ \cite{Mateos:2011tv}. Our finding that $\sigma_z/\sigma_\perp$ is always smaller than one may seem surprising, but we have in fact found an extreme version of this phenomenon in the singular geometry of \cite{Janik:2008tc}. While in this case both DC conductivities vanish, the ratio of AC conductivities $\sigma_z(\omega)/\sigma_\perp(\omega)\to 0$ as $\omega\to 0$ for both oblate and prolate anisotropies (see Fig.~11 of \cite{Rebhan:2011ke}).

Following the procedure of \cite{Kovtun:2003wp}, one can also obtain two related diffusion constants $D_{z,\perp}=\sigma_{z,\perp}\Xi^{-1}$ with common charge susceptibility
\be
\Xi=\left[\int_0^{u_h}du\frac{-g_{\rm eff}^2 g_{tt} g_{uu}}{\sqrt{-g}}\right]^{-1}.
\ee
We have checked \cite{RSprep} 
the results for $D_{z,\perp}$ by numerical calculation of the lowest quasi-normal mode in the diffusive channel, thus verifying that the membrane paradigm is indeed valid 
for conductivities of an anisotropic horizon.

{\it Shear viscosities.} 
In an anisotropic fluid with axial symmetry, 
the viscosity tensor \cite{LL:VII} $\eta_{ijkl}$ has five independent components, two of which are shear viscosities which we denote as
\be
\eta_\perp=\eta_{xyxy},\quad \eta_\parallel=\eta_{xzxz}=\eta_{yzyz}
\ee
and which can be extracted by Kubo formulas from the corresponding retarded correlation functions of the stress tensor. In the gravity dual we need to consider metric fluctuations $h_{\mu\nu}$, for which we adopt the gauge $h_{Mu}=0$.
Expanding the action (\ref{bulkaction}) to second order in $h_{\mu\nu}$ we obtain
\bea
 S^{(2)}&=&\frac{1}{16\pi G}\int d^5 x \Bigl[\sqrt{-g}^{(2)}2A^{(0)}\nonumber\\
&&\qquad+\sqrt{-g}^{(0)}\left(
R^{(2)}-\frac{1}{2}e^{2\phi}a^2g^{zz(2)}\right)\Bigr],
\eea
where we introduced
\be
A^{(0)}=-\frac{1}{2}(8+\frac{1}{2}\phi'^2g^{uu}+\frac{1}{2}e^{2\phi}a^2g^{zz})^{(0)}=\frac{R_{xx}^{(0)}}{g_{xx}^{(0)}}.
\ee

Considering first Fourier modes $\psi_\perp(u,q)\equiv h^x_y(u,q)$ and their conjugate momenta $\Pi_\perp(u,q)$ we find that requiring regularity at the horizon as above for vector perturbations gives the purely transverse shear viscosity
\be\label{etaperp}
\eta_\perp=\frac{\Pi_\perp(u_h,q)}{i\omega\psi_\perp(u_h,q)}
=\frac{\sqrt{\gamma(u_h)}}{16\pi G} =\frac{s}{4\pi}
\ee
which reproduces the universal value for Einstein gravity with isotropic horizons. Because the radial evolution becomes trivial as $q^\mu\to0$, 
\be
\partial_u \Pi_\perp(u,q)=\frac{\sqrt{-g}}{16\pi G}
(g^{tt}\omega^2+g^{zz}q_z^2)\psi_\perp(u,0)\to0,
\ee
the (momentum independent) result of the membrane paradigm gives the purely transverse shear viscosity of the boundary theory at $q^\mu=0$, exactly as in the universality proof of
\cite{Iqbal:2008by}.

\begin{figure}
\includegraphics[width=0.45\textwidth]{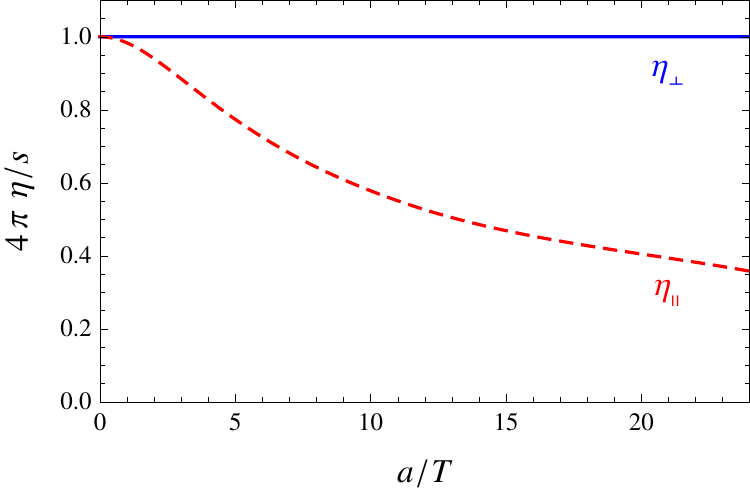}
\caption{Transverse and longitudinal shear viscosities over $s/4\pi$
as a function of the anisotropy parameter
$a/T$.\label{fig:etas}}
\end{figure}

Turning to nontransverse metric fluctuations we find that the
radial flow is still trivial in the limit $q^\mu\to0$ for the case of
$\psi_L=g^{xx}h_{zx}$ because
\be
\partial_u \Pi_L(u,q)=\frac{\sqrt{-g}}{16\pi G}\frac{g_{xx}}{g_{zz}}
(g^{tt}\omega^2+g^{yy}q_y^2)\psi_L(u,0)\to0
\ee
for any $a$ (while $g^{zz}h_{xz}$ would have a nontrivial flow $\propto a^2$).
We can therefore use the membrane paradigm to calculate the
longitudinal shear viscosity from
\bea\label{etalong}
\eta_\parallel&=&\frac{\Pi_\parallel(u_h,q)}{i\omega\psi_\parallel(u_h,q)}
=\frac{\sqrt{\gamma(u_h)}}{16\pi G}\frac{g_{xx}(u_h)}{g_{zz}(u_h)}\nonumber\\
&=&\frac{s}{4\pi\mathcal H(u_h)} < \frac{s}{4\pi}
\eea
for $a\not=0$, which is plotted in Fig.~\ref{fig:etas} as a function of $a/T$.
Note that the ratio $\eta_\parallel/\eta_\perp$ turns out to be identical to $\sigma_z/\sigma_\perp$. 

We have also checked (\ref{etalong}) by a direct numerical evaluation of the Kubo formula
\be
\eta_\parallel=\lim_{\omega\to0}\lim_{u\to0}{\rm Im}\,\frac{\Pi_L(u,\omega,0)}{\omega \psi_L(u,\omega,0)}.
\ee

{\it Momentum diffusion.} 
As a further corroboration, we have 
also studied the diffusive behavior of the relevant shear modes.
The longitudinal shear viscosity is related to the diffusive pole of
the shear mode $Z_3\equiv g^{xx}\big(qh_{tx}(t,z,u)+\omega h_{zx}(t,z,u)\big)$ at small but nonvanishing momenta, where $\vec q$ is oriented in the $z$-direction in order to preserve O(2) symmetry which avoids a coupling to axion modes (the general case will be discussed elsewhere \cite{RSprep}). We can use the trick of \cite{Kovtun:2003wp} of Kaluza-Klein dimensional reduction in the $x$-direction, which maps the problem to that of U(1) gauge fields with a $u$-dependent effective coupling. This yields the diffusion constant
\bea\label{Dzx}
D_{zx}&=&\frac{\eta_\parallel}{\epsilon+P_\perp}\\
&=&\frac{e^{-\frac{3 \phi (u_h)}{4}}}{u_h^3 \sqrt{\mathcal{H}(u_h)}} \int_0^{u_h}du\ u^3
   e^{\frac{3 \phi (u)}{4}} \sqrt{\frac{\mathcal{B}(u)}{\mathcal{H}(u)}}. \nonumber
\eea
We have also checked this result 
by calculating numerically the lowest lying quasinormal mode of the retarded propagator of the shear mode $Z_3(u)$ which satisfies the ordinary differential equation
\bea
 Z_3''+\frac{1}{2} \biggl[\frac{\left(g_{tt}
   g_{zz}'-g_{zz}g_{tt}'\right) \left(q^2
   g_{tt}-\omega ^2 g_{zz}\right)}{g_{tt}
   g_{zz}\left(q^2 g_{tt}+\omega ^2
   g_{zz}\right)} 
-\frac{g_{uu}'}{g_{uu}}&\nonumber\\
+\frac{4 g_{xx}'}{g_{xx}}\biggr]Z_3' 
-g_{uu}\left(\frac{\omega
   ^2}{g_{tt}}+\frac{q^2}{g_{zz}}\right)Z_3=0.\quad&
\eea

We can now compare (\ref{Dzx}) with the analogous calculation of the
shear mode $Z_1\equiv g^{yy}\big(qh_{ty}(t,x,u)+\omega h_{ty}(t,x,u)\big)$
with momentum oriented in the $x$ direction, yielding
(the breaking of SO(2) invariance in this case does not lead to
complications from axion modes)
\bea\label{Dxy}
D_{xy}
&=&\frac{e^{-\frac{3 \phi (u_h)}{4}} \sqrt{\mathcal{H}(u_h)}}{u_h^3} \int_0^{u_h}du\ u^3
   e^{\frac{3 \phi (u)}{4}} \sqrt{\frac{\mathcal{B}(u)}{\mathcal{H}(u)}}\nonumber\\
&=&\frac{\eta_\perp}{\epsilon+P_\perp}={1\04\pi T}
\eea
where the last equality follows from (\ref{etaperp})
together with (\ref{entropy}), but is also reproduced numerically.
Taking the ratio of Eqs.~(\ref{Dzx}) and (\ref{Dxy}) 
leads again to the 
result
\be\label{etapa}
\frac{\eta_\parallel}{s}=\frac{1}{4\pi \mathcal H(u_h)}
<\frac{1}{4\pi} \quad \mbox{for $a\not=0$}.
\ee

{\it Conclusion.} As mentioned above, a result for $\eta_\parallel$ that deviates from the universal result of Einstein gravity with isotropic horizons has been recently obtained in a dual model for p-wave superfluids \cite{Erdmenger:2010xm
}, however one that is above $s/4\pi$. To our knowledge, the result (\ref{etapa}) is the first example of a shear viscosity that falls below this value without recourse to higher-derivative gravity and moreover with fully known gauge-gravity correspondence. 

Parts of the phase diagram in $a/\mu$ and $T/\mu$ of this theory exhibit instabilities against filamentation into inhomogeneous mixed configurations with $a_{\rm iso}=0$ and some $a_{\rm ani}>a$, while for sufficiently large $a$ the anisotropic homogeneous phase is thermodynamically stable \cite{Mateos:2011tv}. Strictly speaking, the above calculation of hydrodynamical transport coefficients is limited to stable homogeneous phases, but we have numerically verified their existence on lines of constant $a/T \gtrapprox 1.3$.

It is not clear what implications, if any, our result of $\eta_\parallel/\eta_\perp<1$ for a strongly coupled stationary anisotropic plasma would have for real nonequilibrium evolution.
Intrinsically anisotropic hydrodynamic descriptions of quark-gluon plasma have recently been proposed in \cite{Florkowski:2010cf
}, however without allowing for different shear viscosity components. It would be interesting to study the effects of different shear viscosities in a hydrodynamical evolution, and also what ratios $\eta_\parallel/\eta_\perp$ a weakly coupled anisotropic non-Abelian plasma would give rise to.

{\it Acknowledgments.} We would like to thank Alex Buchel, Johanna Erdmenger, Patrick Kerner, Andreas Schmitt, Aleksi Vuorinen, and Hansj\"org Zeller
for discussions. This work was supported by the Austrian Science
Fund FWF, project no. P22114.

\bibliographystyle{prsty}
\bibliography{ar,tft,qft,books}
\end{document}